\documentclass[epj]{svjour}
\usepackage{bm}
\usepackage{epsfig,graphics}
\usepackage{amsfonts,amssymb,stmaryrd,latexsym,amsmath}
\usepackage{slashed}

\newcommand{\be}{\begin{equation}}
\newcommand{\ee}{\end{equation}}
\newcommand{\ba}{\begin{eqnarray}}
\newcommand{\ea}{\end{eqnarray}}
\newcommand{\non}{\nonumber}
\newcommand{\Lag}{\mathcal{L}}
\newcommand{\Amp}{\mathcal{A}}
\newcommand{\Eng}{\mathcal{E}}
\newcommand{\lang}{\left\langle}
\newcommand{\rang}{\right\rangle}

\begin{document}

\title{Bound state nature of the exotic $\bm{ Z_b}$ states}
\author{Martin Cleven\inst{1}\thanks{{\it Email address:} m.cleven@fz-juelich.de},
        Feng-Kun Guo\inst{2}\thanks{{\it Email address:} fkguo@hiskp.uni-bonn.de},
        Christoph Hanhart\inst{1}\thanks{{\it Email address:} c.hanhart@fz-juelich.de},
        and Ulf-G. Mei{\ss}ner\inst{1,2}\thanks{{\it Email address:} meissner@hiskp.uni-bonn.de}}
\institute{Institute for Advanced Simulation and J\"ulich Center for Hadron
          Physics, Institut f\"{u}r Kernphysik, Forschungszentrum J\"{u}lich, D-52425 J\"{u}lich,
          Germany \and
          Helmholtz-Institut f\"ur Strahlen- und Kernphysik and
          Bethe Center for Theoretical Physics, Universit\"at Bonn, D-53115 Bonn, Germany}

\date{Received: date / Revised version: date}
%\maketitle

\abstract{ The assumption that the newly observed charged bottomonia states
$Z_b(10610)$ and $Z_b(10650)$ are of molecular nature is confronted with the
measured invariant mass distributions for the transitions of the $\Upsilon(5S)$
to the final states $h_b\pi^+\pi^-$ and $h_b(2P)\pi^+\pi^-$. It is shown that
the assumption that the $Z_b(10610)$ and $Z_b(10650)$ are $B\bar B^*+{\rm c.c.}$
and $B^*\bar B^*$ bound states, respectively,  with very small binding energies
is consistent with the data. The calculation is based on a power counting for
bottom meson loops, which is explicitly given up to two-loop in the framework of
a nonrelativistic effective field theory. We also show that if the $Z_b$ states
are of molecular nature, then the data should not be analyzed by using a
Breit-Wigner parametrization.
 }


\authorrunning{M.~Cleven {\it et al.}}
\titlerunning{Bound state nature of the $Z_b$ states}

%\vspace{1cm}
%
%\newpage

\maketitle

\medskip

\section{Introduction}

Very recently, the Belle Collaboration reported the observation of two charged
bottomonia states $Z_b(10610)$ and $Z_b(10650)$ in five different decay channels
of the $\Upsilon(5S)$~\cite{Adachi:2011gj}. Their masses and widths from
averaging the measurements in various channels are $M_{Z_b}=10608.4\pm2.0$~MeV,
$\Gamma_{Z_b}=15.6\pm2.5$~MeV, and $M_{Z_b'}=10653.2\pm1.5$~MeV,
$\Gamma_{Z_b'}=14.4\pm3.2$~MeV, respectively.

The lower $Z_b$ lies very close to the $B\bar B^*$ threshold, $10604$~MeV, and
the higher one is close to the $B^*\bar B^*$ threshold, $10650$~MeV. The
reported masses are slightly above the corresponding thresholds. Very soon after
the discovery, it was proposed that the two $Z_b$ states are of molecular
nature~\cite{Bondar:2011ev}. To be precise, the main components in the wave
functions of the $Z_b(10610)$ and $Z_b(10650)$ (to be called $Z_b$ and $Z_b'$)
are $B\bar B^*+{\rm c.c.}$ and $B^*\bar B^*$, respectively. Much attention has
been paid to this molecular interpretation. Calculations from QCD sum
rules~\cite{Zhang:2011jj}, constituent quark models~\cite{Yang:2011rp} and
one-boson-exchange model of bottom meson potentials~\cite{Sun:2011uh} claimed
the existence of $I=1,J^{PC}=1^{+-}$ $B\bar B^*$ and $B^*\bar B^*$ molecular
states, and obtained masses consistent with the measured values for the two
$Z_b$ states. In Ref.~\cite{Nieves:2011vw}, the authors argue that the
one-pion-exchange potential does not support an $S$-wave $B\bar B^*$ resonance
state above threshold based on an effective field theory. However, as will be
shown in this paper, the experimental data in the $h_b\pi^+\pi^-$ and
$h_b(2P)\pi^+\pi^-$ channels are also consistent with masses slightly lower than
the corresponding thresholds.

As already known from many other processes and for a long time, coupled channels
can produce peaks at their thresholds as a result of the unitary cut. This kind
of effect was also noticed for the case of the $Z_b$ states
in~\cite{Bugg:2011jr,Danilkin:2011sh,Chen:2011pv}.

In fact, the existence of an isospin vector exotic state with $J^P=1^+$ at the
bottomonium mass region was proposed many years ago~\cite{Voloshin:1982ij}.
There is a long-standing puzzle in the decay $\Upsilon(3S)\to
\Upsilon(1S)\pi\pi$. Before the measurements of the $\pi\pi$ invariant mass
spectra of the transitions from the $\Upsilon(4S)$ to the lower-lying $\Upsilon$
states with the emission of two pions, two evident bumps show up in the $\pi\pi$
invariant mass spectrum of the decay $\Upsilon(3S)\to \Upsilon(1S)\pi\pi$, see
e.g.~\cite{Butler:1993rq}. This peculiar structure is quite different from that
of the transitions $\psi(2S)\to J/\psi\pi\pi$ and $\Upsilon(2S)\to
\Upsilon(1S)\pi\pi$. Many models were proposed in order to resolve this puzzle.
It was shown that by including the mentioned hypothetical exotic state as well
as  the $\pi\pi$ final state interaction, one can describe all the data of the
$\Upsilon(nS)\to\Upsilon(mS)\pi\pi$ ($n>m,n=2,3$) transitions
well~\cite{Anisovich:1995zu,Guo:2004dt}. Later on, it was shown that the
double-bump structure of the dipion invariant mass spectra of the transitions
$\Upsilon(4S)\to\Upsilon(1S,2S)\pi\pi$ can also be described with the same
exotic particle~\cite{Guo:2006ai}. However, in all these analyses only one state
was included, while the Belle Collaboration reported two $Z_b$ states. Very
recently, after the discovery of the $Z_b$ states, they were shown to play an
important role in the helicity angular distribution of the transition
$\Upsilon(5S)\to \Upsilon(2S)\pi\pi$~\cite{Chen:2011zv}. In all the above
studies, effects of bottom and anti-bottom meson loops which can couple to the
exotic states were not taken into account. In view of the fact that these two
$Z_b$ states are in the vicinities of the $B\bar B^*$ and $B^*\bar B^*$
thresholds, respectively, the bottom meson loops could affect the line shapes of
the $Z_b$ states significantly analogous to the $K\bar K$ loop effects on the
$f_0(980)$ and $a_0(980)$~\cite{Flatte:1976xu}. A systematic (re)analysis of all
the $\Upsilon(nS)\to\Upsilon(mS)\pi\pi$ ($n>m,n=2,3,4,5$) including both $Z_b$
states, bottom and anti-bottom meson loops and $\pi\pi$ final state interaction
is necessary~\cite{future}. In this paper, we will however focus on a simpler
task, namely study the $Z_b$ states in the decays of $\Upsilon(5S)$ to
$h_b\pi^+\pi^-$ and $h_b(2P)\pi^+\pi^-$ which are dominated by the $Z_b$ states
as indicated by the data~\cite{Adachi:2011gj}, and check if the molecular
assumption is consistent with the available experimental information. Especially
we demonstrate below that the data allow for bound state poles located below the
$\bar BB^*$ and $\bar B^*B^*$ thresholds, respectively.

\section{Lagrangians}
\label{sec:Lag}

Because the spin-dependent interaction between a heavy quark and a gluon is
suppressed by $1/m_Q$, with $m_Q$ the heavy quark mass, in the heavy quark
limit, the spin of heavy quarks decouples. Hence, it is convenient to introduce
heavy hadrons and heavy quarkonia in terms of spin multiplets. In the rest frame
$v^\mu=(1,\vec{0})$, $v$ being the heavy quark velocity, one has
$J=\vec{\Upsilon}\cdot\vec{\sigma}+\eta_b$ with $\vec{\Upsilon}$ and $\eta_b$
annihilating the $\Upsilon$ and $h_b$, respectively, and
$H_a=\vec{V}_a\cdot\vec{\sigma}+P_a$, with $\vec{V}_a$ and $P_a$ annihilating
the vector and pseudoscalar heavy mesons, respectively. $\sigma^i$ are the Pauli
matrices, and $a$ is the light flavor index. Explicitly, one can write
$P_a(V_a)=\left(B^{(*)-},\bar B^{(*)0}\right)$ for bottom mesons. The heavy
mesons containing an anti-heavy quark are collected in $\bar H_a=-\vec{\bar
V}_a\cdot\vec{\sigma}+\bar P_a$~\cite{Fleming:2008yn}.

The Lagrangian for the coupling of the $\Upsilon$ to the bottom and anti-bottom
mesons can be obtained by evaluating the trace in the Lagrangian
\be%
\Lag_{\Upsilon} =  i \frac{g_2}{2} \lang J^\dag H_a \vec{\sigma}\cdot
\!\overleftrightarrow{\partial}\!{\bar H}_a\rang + {\rm H.c.}, \label{eq:LUps}
\ee%
where $A\overleftrightarrow{\partial}\!B\equiv
A(\vec{\partial}B)-(\vec{\partial}A)B$. The Lagrangian for the coupling the
bottom mesons to the $P$-wave bottomonium $h_b$ reads~\cite{Guo:2010ak}
\be%
{\cal L}_{h_b} = - g_1\epsilon^{ijk}h_b^{\dag i}V_a^j{\bar V}_a^k + ig_1h_b^{\dag
i} \left( V_a^i{\bar P}_a - P_a{\bar V}_a^i \right) + {\rm H.c.}
\ee%

Since the quantum numbers $I^G (J^{P})=1^+(1^{+})$ are favored for both states
by the experimental analysis~\cite{Adachi:2011gj}, the $C$ parity of the neutral
$Z_b$ states should be negative. This means under parity and charge conjugation,
the fields annihilating the $Z_b$ states should transform as
\be%
Z^i \overset{\mathcal P}{\rightarrow} Z^i, \qquad Z^i \overset{\mathcal
C}{\rightarrow} -Z^{iT},
\ee%
and the heavy quark spin symmetry transformation is given by $Z^i
\overset{\mathcal S}{\rightarrow} S Z^i{\bar S}^\dag$ with $S$ and $\bar S$
acting on the bottom and anti-bottom quark fields, respectively. With these
transformation properties, one can construct the Lagrangi\-an for an $S$ wave
coupling of the $Z_b$ states to the bottom and anti-bottom mesons,
\be%
\Lag_Z = i \frac{z^{\rm bare}}{2} \lang Z^{\dag i}_{ba} H_a \sigma^i{\bar
H}_b\rang + {\rm H.c.}, \label{eq:Lz}
\ee%
with $z^{\rm bare}$ is the bare coupling constant, which will be renormalized to
the physical one $z=z^{\rm bare}\sqrt{Z}$ where $Z$ is the wave function
renormalization constant. %, see Sec.~\ref{sec:fit}.
The three different charged states are collected in a $2\times2$ matrix as
$$ Z^i_{ba} = \left(
         \begin{array}{cc}
           \frac1{\sqrt{2}} Z^{0i} & Z^{+i} \\
           Z^{-i} & - \frac1{\sqrt{2}} Z^{0i} \\
         \end{array}
       \right)_{ba}.
$$

The axial coupling of the pion fields to the heavy and anti-heavy mesons in
heavy flavor chiral perturbation theory at the lowest order is given
by~\cite{Burdman:1992gh,Fleming:2008yn}
\be%
\Lag_\pi = \frac{g}{\sqrt{2}F_\pi} \lang H^\dag_{a} H_b
\vec{\sigma}\cdot\vec{\nabla}\phi_{ba}\rang - \frac{g}{\sqrt{2}F_\pi} \lang \bar
H^\dag_{a}\vec{\sigma}\cdot\vec{\nabla}\phi_{ab} \bar H_b \rang . \label{eq:Lpi}
\ee%

\section{Propagator of the $Z_b$ states}
\label{sec:prop}

\begin{figure}[t]
\begin{center}
\includegraphics[width=0.48\textwidth]{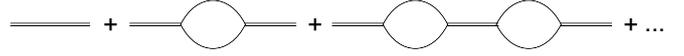}
\caption{Expansion of the two-point Green's function. Doubles lines and bubbles
represent the bare propagators and self-energies, respectively. \label{fig:prop}}
\end{center}
\end{figure}

The propagator of the $Z_b$ states is given by the two-point Green's function
\be%
\delta^{ij} \delta^{ab} G_Z(E) \equiv \int d^4x e^{-iEt}\lang 0\left| T\{
Z^{i}_a(x) Z^{\dag j}_b(0) \} \right|0\rang,
\ee%
where $i,j$ and $a,b$ and the indices for spin and isospin, respectively, and
$T$ denotes time order. Figure~\ref{fig:prop} illustrates the renormalization of
the Green's function to one loop order.
The bare propagator $i/[2(E-\Eng_0)]$ is dressed by the self-energy $-i\Sigma$.
Therefore, the full propagator can be written as
\ba%
G_Z(E) = \frac12 \frac{i}{E-\Eng_0-\Sigma(E)}.
\ea%
Our convention is such that the non-relativistic normalization differs from the
relativistic one by a factor of $1/\sqrt{M_Z}$. The self-energy in
$d$-dimensional space-time reads
\ba%
\Sigma(E) &=& i\frac{(z^{\rm bare})^2}{4} \int\!\! \frac{d^dl}{(2\pi)^d}\,
\frac1{l^0-\vec{l}^2/(2m_1)+i\epsilon} \non\\
&&\times \frac1{E-l^0-\vec{l}^2/(2m_2)+i\epsilon}.\label{eq:sigma0}
\ea%
Notice that a factor of 2 has been multiplied in this definition in order to
take into account both $B\bar B^*$ and its charge conjugated channel. The same
factor appears in the $B^*\bar B^*$ self-energy due to a different reason:
$\epsilon_{ijk}\epsilon_{i'jk}=2\delta^{ii'}$. In dimensional regularization
with the $\overline{\rm MS}$ subtraction scheme, this integration is finite for
$d=4$. It corresponds to an implicit subtraction of the linear divergence which
appears at $d=3$. Taking $d=4$, one has
\ba%
\Sigma^{\overline{\rm MS}}(E) &=& (z^{\rm bare})^2 \frac{\mu}{8\pi} \sqrt{-2\mu
E-i\epsilon} \non\\
&=& (z^{\rm bare})^2 \frac{\mu}{8\pi} \left[\sqrt{{-}2\mu E}\theta({-}E) {-} i
\sqrt{2\mu E}\theta(E) \right], \non\\ \label{eq:sigma}
\ea%
where $\mu=m_1m_2/(m_1+m_2)$ is the reduced mass, and $\theta(E)=1$ for positive
$E$ and 0 for negative $E$ is the step function. The superscript $\overline{\rm
MS}$ denotes the subtraction scheme.

The bare energy $\Eng_0$ is renormalized to the physical energy, $\Eng$, at the
mass of the $Z_b$. $\Eng$ is connected to the mass of the $Z_b$ as
$\Eng=M_Z-m_1-m_2$ with $m_{1,2}$ being the masses of mesons in the loop. The
renormalization condition is such that $\Eng$ is a zero of the real part of the
denominator of the propagator. Thus, $\Eng=\Eng_0+{\rm Re}\Sigma(\Eng)$, and
expanding the real part of
 the self-energy $\Sigma(E)$ around $E=\Eng$
gives
\ba%
G_Z(E) &=& \frac12\frac{i}{(E-\Eng)[1-{\rm Re}(\Sigma'(\Eng))]-\widetilde{\Sigma}(E)} \non\\ &=& \frac12
\frac{iZ}{E-\Eng-Z\widetilde{\Sigma}(E)} , \label{eq:gz0}
\ea%
where the wave function renormalization constant is
$Z=\left[1-{\rm Re}(\Sigma'(\Eng))\right]^{-1}$ with $\Sigma'(\Eng)$ representing the
derivative of $\Sigma(E)$ with respect to $E$ at $E=\Eng$, and
$$\widetilde{\Sigma}(E)=\Sigma(E)-{\rm Re}(\Sigma(\Eng))-(E-\Eng){\rm Re}(\Sigma'(\Eng)) \ .$$
Particularly, ${\rm Re}(\widetilde{\Sigma}(\Eng))={\rm Re}(\widetilde{\Sigma}'(\Eng))=0$.
In the standard scenario (absence of nearby thresholds; stable states)
$\widetilde{\Sigma}(E)$ is dropped, however, here, due to the very close
branch point singularity at $E=0$, this function not only
acquires an imaginary part for $E>0$ but also  varies rapidly. It therefore needs to be kept in the
propagator.
 Eq.~(\ref{eq:gz0}) is valid for both $\Eng>0$
as well as $\Eng<0$ (but ill defined for $\Eng=0$). The expression for $Z$
follows from Eq.~(\ref{eq:sigma}),
\be%
Z = \left[ 1+\frac{\mu^2(z^{\rm bare})^2}{8\pi\gamma} \right]^{-1}
\theta(-\Eng) + 1\times \theta(\Eng).
\ee%

If the $Z_b$ ($Z_b'$) is a pure $B\bar B^*$ ($B^*\bar B^*$) bound state, the
wave function renormalization constant should be 0 since $1-Z$ measures the
probability of finding a bound state in the physical state, see e.g.
Refs.~\cite{molecule1}. This means the bare coupling $z^{\rm bare}$ goes to
infinity. However, the physical effective coupling is finite
\be%
(z^{\rm eff})^2 = \lim_{|z|\to\infty} Z (z^{\rm bare})^2 =
\frac{8\pi}{\mu^2}\gamma \ , \label{eq:zeff}
\ee%
with the binding momentum $\gamma = \sqrt{-2\mu \Eng}$. Eq.~(\ref{eq:zeff})
coincides with the one for an $S$ wave loosely bound state derived in
Refs.~\cite{molecule1} taking into account the factor $2$ as discussed below
Eq.~(\ref{eq:sigma0}).

Furthermore, the $Z_b$ states can also decay into channels other than the bottom
and anti-bottom mesons, such as $\Upsilon(nS)\pi~(n=1,2,3)$, $h_b(1P,2P)\pi$,
$\eta_b\rho$ and so on. For the complete propagator we therefore need to write
\ba%
G_Z(E) = \frac12
\frac{iZ}{E-\Eng-Z\widetilde{\Sigma}(E) + i
\Gamma^{\rm phys}(E)/2}.
\ea%

According to the power counting in the non-relativistic effective field theory
analyzed in details in Ref.~\cite{Guo:2010ak}, the transition amplitude between
the $Z_b$ states, which couple to the bottom mesons in an $S$-wave, and $S$-wave
bottomonia should scale as $q^2/(M_B^2v_B)$, with $q$ being the external
momentum, and $v_B$ the velocity of the intermediate bottom mesons. Because
$q\ll M_Bv_B^{1/2}$, this is a suppression factor. On the other hand, the
transition amplitudes to $P$-wave bottomonia $h_b(1P,2P)$ scales as $q/v_B$.
Since $v_B\ll1$, one would expect the decays from the $Z_b$ to $h_b(1P,2P)\pi$
dominate those to $\Upsilon(nS)\pi$. Therefore we assume that $\Gamma^{\rm
phys}$ is saturated by the former channels.

\subsection{Power counting of two-loop diagrams}
\label{sec:twoloop}

\begin{figure}[t]
\begin{center}
\includegraphics[width=0.48\textwidth]{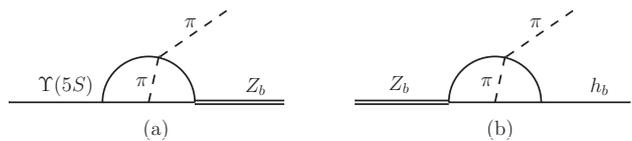}
\caption{Two loop diagrams for the subprocesses $\Upsilon(5S)\to Z_b\pi$ (a) and $Z_b\to h_b\pi$ (b). Solid lines in the loops
represent bottom and anti-bottom mesons. \label{fig:twoloop}}
\end{center}
\end{figure}

So far we have only considered one-loop diagrams. There can be more loops by
exchanging pions between (anti-) bottom mesons. Following the formalism set up
in Refs.~\cite{Guo:2010ak,Guo:2009wr,Guo:2010zk}, we can analyze the power
counting of these higher order loops. For processes with intermediate heavy
meson loops, if the virtuality of these intermediate heavy mesons is not large,
their three-momenta are small compared with their masses. Hence these heavy
mesons can be dealt with nonrelativistically, and one can set up a power
counting in terms of the velocity of the heavy mesons, $v_B$. In this power
counting, the momentum and energy of the intermediate mesons scale  as $v_B$ and
$v_B^2$, respectively, and hence the measure of one-loop integration scales as
$\int d^4l\sim v_B^5$.

There are two different topologies to be distinguished. On the one hand there
are vertex corrections --- those are diagrams where either one or more pions are
exchanged or where a four--$B$-meson contact operator is inserted, followed by a
two--heavy meson propagator. Since the typical momentum in the loop is a lot
larger than the pion mass and the pion couples with a $P$-wave, the vertices for
the exchanged pions as well as the short ranged operator provides a factor
$v_B^\lambda$, with $\lambda\geq 0$, to the power counting of the diagram. In
addition, the two-loop vertex correction has one more integral measure as well
as two more propagators, each $\sim v_B^{-2}$. Thus in total a vertex correction
appears to be suppressed by a factor
$v_B^{5+\lambda}/(v_B^2)^2=v_B^{1+\lambda}$. However, one needs to pay special
attention to the two-loop diagrams of the kind shown in Fig.~\ref{fig:twoloop},
where, e.g., a pion gets produced on one heavy meson and rescatters off the
other one before going on-shell. These diagrams need to be analyzed case by
case.

Let us first analyze the power counting of the diagram Fig.~\ref{fig:twoloop}(a)
for $\Upsilon(5S)\to Z_b\pi$.  The leading order amplitude for the bottom
meson--pion scattering formally scales as $(E_{\pi1}+E_{\pi2})/F_\pi^2$, see
e.g.~\cite{Guo:2009ct}, with $E_{\pi1,2}$ the energies of the two pions.  Due to
some subtle cancellation mechanisms also the energy of the exchanged pion gets
put on-shell in this vertex --- in analogy to what happens in the reaction
$NN\to NN\pi$~\cite{nn2dpi}.  For the numerical estimates below we use
$E_\pi=M_{\Upsilon (5S)}-M_{Z_b^{(\prime)}}\simeq 250$ MeV.  There are two
$P$-wave couplings in the two-loop diagrams: the coupling of the $\Upsilon(5S)$
to the bottom and anti-bottom mesons and the vertex emitting a pion inside the
loop.  The $Z_b$ bottom meson vertex is in an $S$-wave. The momenta from the two
$P$-wave vertices can contract with each other, and hence scale as $v_B^2$
(recall in the one-loop case, the $\Upsilon(5S)B\bar B$ $P$-wave vertex must
contract with the external momentum, and hence scales as $q$~\cite{Guo:2010ak}).
There are five propagators, and each of them has a contribution of order
$v_B^{-2}$. Therefore, the power counting of the two-loop diagram of
Fig.~\ref{fig:twoloop}(a) reads \be \frac{(v_B^5)^2 v_B^2}{(v_B^2)^5}
\frac{E_\pi}{16\pi^2F_\pi^2} M_B = \frac{v_B^2E_\pi
  M_B}{\Lambda_\chi^2}, \label{eq:pctwoloop1} \ee
where the factor $1/16\pi^2$ appears because there is one more loop compared to the
one-loop case, and the chiral symmetry breaking scale is denoted as
$\Lambda_\chi=4\pi F_\pi$. We have introduced a factor of $M_B$ to make the whole
scaling dimensionless. One may estimate $v_B\sim
\sqrt{(\hat{M}-2\hat{M}_B)/\hat{M}_B}\simeq 0.15$ with
$\hat{M}=(M_{\Upsilon(5S)}+M_Z)/2$ and $\hat{M}_B$ the average mass of the bottom
mesons $B$ and $B^*$. This is to be compared to the one-loop diagram, which scale
as $q^2/(M_B^2v_B)$ --- see Refs.~\cite{Guo:2010zk,Guo:2010ak} and the previous
section. Numerically, it is of similar size as or even smaller than the two-loop
diagram given in Eq.~(\ref{eq:pctwoloop1}). Hence, the two-loop diagram could be
more important than the one-loop diagrams. However, the same mechanisms that
suppress vertex corrections should also suppress three-- or more--loop diagrams.

The situation for the transition $Z_b\to h_b\pi$ is different. The two-loop
diagram is shown in Fig.~\ref{fig:twoloop}(b). Now there is only one $P$-wave
vertex, which is the bottom meson--pion vertex. All the other three vertices are
in an $S$-wave. Due to the $P$-wave nature of the decay $Z_b\to h_b\pi$, the
amplitude must be proportional to the external momentum. Therefore, the only
$P$-wave vertex should scale as $q$, and the product of the other three vertices
scales, again, as $E_{\pi}$. Taking into account further the loop integral
measure and the propagators, the power counting for this diagram reads
\be%
\frac{(v_B^5)^2}{(v_B^2)^5} \frac{E_\pi}{16\pi^2F_\pi^2} q M_B =
q\frac{M_B E_\pi}{\Lambda_\chi^2}, \label{eq:pctwoloop2}
\ee%
where $M_B$ is again introduced to render the scaling dimensionless. The
one-loop diagrams scale like the transitions between two $P$-wave heavy
quarkonia. According to Ref.~\cite{Guo:2010ak}, the power counting is given by
$q/v_B$, which is numerically much larger than the scaling for the two-loop
diagram given in Eq.~(\ref{eq:pctwoloop2}). Hence, the two-loop diagrams can be
safely neglected for the $Z_b\to h_b\pi$.

In addition, counter-terms of the kind $\Upsilon Z_b\pi$ need to be considered
for they are needed to absorb the divergencies of the loop diagrams just
discussed and are expected to be of the same importance as the loops. The
corresponding Lagrangian to leading order of the chiral expansion reads
\be%
\Lag_{\Upsilon Z_b\pi} = c \Upsilon^i Z_{ba}^{\dag i} \partial^0\phi_{ab} + {\rm
H.c.} \label{eq:Lconstant}
\ee%
 The analogous counter-terms for $Z_b h_b \pi$ can
be dropped here for they are suppressed compared to the one loop diagram.
\begin{figure}[t]
\begin{center}
\includegraphics[height=2.5cm]{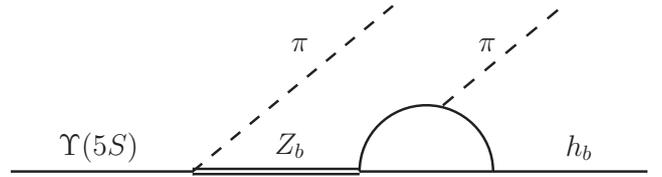}
\caption{Decay mechanism of the process $\Upsilon(5S)\to Z_b\pi\to h_b\pi\pi$. Solid lines in the loop
represent bottom and anti-bottom mesons. \label{fig:oneloop}}
\end{center}
\end{figure}
Assuming the $Z_b$ and $Z_b'$ are spin partners of each
other~\cite{Bondar:2011ev}, one can use the same coupling constant for them.

A full analysis would not only require the evaluation of all the diagrams
mentioned above but also those where there are no $Z_b^{(\prime)}$ present in
the processes. This will be studied in a later publication~\cite{future}. Here
we take a more pragmatic point of view and simply represent the whole $\Upsilon
Z_b\pi$ transition by the single contact term of Eq.~(\ref{eq:Lconstant}). The
full transition is illustrated in Fig.~\ref{fig:oneloop}.

One comment is in order: the Lagrangian of Eq.~(\ref{eq:Lconstant}) could as
well mimic a compact component of the $Z_b^{(\prime)}$. This is important
especially because both the two-loop and one-loop scaling given by
Eq.~(\ref{eq:pctwoloop1}) and $q^2/(M_B^2 v_B)$ are much smaller than 1, and
hence a small compact component could be more important than the bottom meson
loops in the $\Upsilon Z_b\pi$ vertex. Thus, since we expect this contact term
to appear at leading order, it seems as if we would not be able to disentangle a
compact, say, tetraquark component from a molecular one. However, since in the
molecular scenario the transition $Z_b\to \pi h_b$ is dominated by the loop, the
structure can indeed be tested, since for molecules the dynamics appears to be
quite restricted. Thus, in the present approach the $\Upsilon (5S)\pi$ vertex
provides a source term for the $Z_b^{(\prime)}$, while their decays are the
subject of this study.

\section{Results}
\label{sec:constaint}

The relevant vertices follow from these Lagrangians. The decay amplitudes for
the two-body transitions can be found in Appendix~\ref{app:amp}. On one hand,
the $\Upsilon(5S)$ is only about 120 and 70~MeV above the $Z_b\pi$ and $Z_b'\pi$
thresholds, respectively. On the other hand, it has a large width
$110\pm13$~MeV, and the experimental data were taken in an energy range around
the $\Upsilon(5S)$ mass. Therefore, when calculating its decay widths, one has
to take into account the mass distribution of the $\Upsilon(5S)$. Its three-body
decay width can then be calculated using
\ba%
\Gamma(\Upsilon(5S))_{\rm 3-body} &=&
\frac1{W}\int_{(M_\Upsilon-2\Gamma_\Upsilon)^2}^{(M_\Upsilon+2\Gamma_\Upsilon)^2}
\!\!ds\, \frac{(2\pi)^4}{2\sqrt{s}} \int d\Phi_3 |\Amp|^2 \non\\
&&\times \frac1{\pi}{\rm
Im}\left(\frac{-1}{s-M_\Upsilon^2+iM_\Upsilon\Gamma_\Upsilon}\right) ,
\ea%
where $M_\Upsilon=10.865$~GeV and $\Gamma_\Upsilon=0.11$~GeV are the mass and
width of the $\Upsilon(5S)$, respectively, and $\int d\Phi_3$ denotes the
three-body phase space, see e.g.~\cite{PDG2010}. The function $\Amp$ contains
all the physics and can be easily obtained using the loop amplitudes given
explicitly in the appendix. Both positively and negatively charged $Z_b$ and
$Z_b'$ states should be considered. The factor $1/W$ with
$$W = \int_{(M_\Upsilon-2\Gamma_\Upsilon)^2}^{(M_\Upsilon+2\Gamma_\Upsilon)^2}
\!\!ds\, \frac1{\pi}{\rm
Im}\left(\frac{-1}{s-M_\Upsilon^2+iM_\Upsilon\Gamma_\Upsilon}\right)$$ is
considered in order to normalize the spectral function of the $\Upsilon(5S)$.

We consider the case that the $Z_b$ and $Z_b'$ couple only to the $B\bar
B^*+{\rm c.c.}$ and $B^*\bar B^*$ channels, respectively. Coupled channel
effects should be suppressed because $|\Eng| \ll M_{B^*}-M_B$ for both $Z_b$
states.

The parameters of the model are the normalization factors $N$, chosen
individually for the two final states, the physical couplings, which are
products of $\sqrt{Z}$ and the bare couplings, for the $Z_b$ states to the
relevant open bottom channels, $z_1$ and $z_2$ (or equivalently the binding
energies of the $Z_b^{(\prime)}$), and those couplings for the $h_b$ and
$h_b(2P)$, denoted by $g_1$ and $g_1'$, respectively. The parameter $c$ of
Eq.~(\ref{eq:Lconstant}) is absorbed into the overall normalization factor. Both
$g_1$ and $g_1'$ only appear in a product with the $z_i$. In order to reduce the
number of free parameters, we assume $g_1'=g_1$ in the following. Note that
neither of them can be measured directly, since the masses of $h_b$ and
$h_b(2P)$ are below the $\bar B B^*$ threshold. In the actual fit we will adjust
$z_1$, $r_z\equiv z_2/z_1$, $g_1z_1$ and the two normalization constants.

Using the amplitudes of Eqs.~(\ref{Aeq:ahb1},\ref{Aeq:ahb2}), we fit
 the parameters to the  invariant mass spectra of both $h_b\pi^+$ and
$h_b(2P)\pi^+$  from 10.56 GeV to 10.70~GeV in the
missing mass spectrum $MM(\pi)$. In the chosen region, there are 14 data points
for the $\Upsilon(5S)\to h_b\pi^+\pi^-$ and 13 for the $\Upsilon(5S)\to
h_b(2P)\pi^+\pi^-$.

The decay widths of the $Z_b$ and $Z_b'$ into $h_b\pi$ are obtained to be
\ba%
\Gamma(Z_b\to h_b\pi) &=& 4.8
\left(\frac{gg_1z_1}{F_\pi} {\rm GeV}^2 \right)^2~{\rm MeV} \non\\
&=& 140 (g_1z_1 {\rm GeV})^2~{\rm MeV}, \non\\
\Gamma(Z_b^{\prime}\to h_b\pi) &=& 5.8
\left(\frac{gg_1z_2}{F_\pi} {\rm GeV}^2 \right)^2~{\rm MeV} \non\\
&=& 169  (g_1z_2 {\rm GeV})^2~{\rm MeV}. \label{eq:widhb}
\ea%
Due to smaller phase space, the widths for the decays $Z_b^{(\prime)}\to
h_b(2P)\pi$ get smaller numerical factors. We find
\ba%
\Gamma(Z_b\to h_b(2P)\pi) = 30 (g_1'z_1 {\rm GeV})^2~{\rm MeV}, \non\\
\Gamma(Z_b^{\prime}\to h_b(2P)\pi) = 46  (g_1'z_2 {\rm GeV})^2~{\rm MeV}.
\label{eq:widhbp}
\ea%
In getting the above numbers, we use $F_\pi=92.4$~MeV.  Because the $B^*$ is
below the $B\pi$ threshold, the axial coupling constant $g$ cannot be directly
measured. Fortunately, there have been quite a few theoretical determinations
using different methods. For a collection of these results,
see~\cite{Li:2010rh}. Almost all the determinations fall in the range between
0.3 and 0.7, and $g=0.5$ is used here.

If the $Z_b$ and $Z_b'$ states are $S$-wave bound states of the $B\bar B^*$ and
$B^*\bar B^*$, respectively, their coupling strengths to the bottom and
anti-bottom mesons are related to the binding energies. The relation has been
derived in Eq.~(\ref{eq:zeff}). The coupling constants $z_i$ appear in
Eqs.~(\ref{eq:widhb}), which enter the $Z_b$ propagators, as well as in the
transition amplitudes --- c.f. Eqs.~(\ref{Aeq:ahb1},\ref{Aeq:ahb2}).

\begin{figure}[t]
\begin{center}
\includegraphics[width=0.48\textwidth]{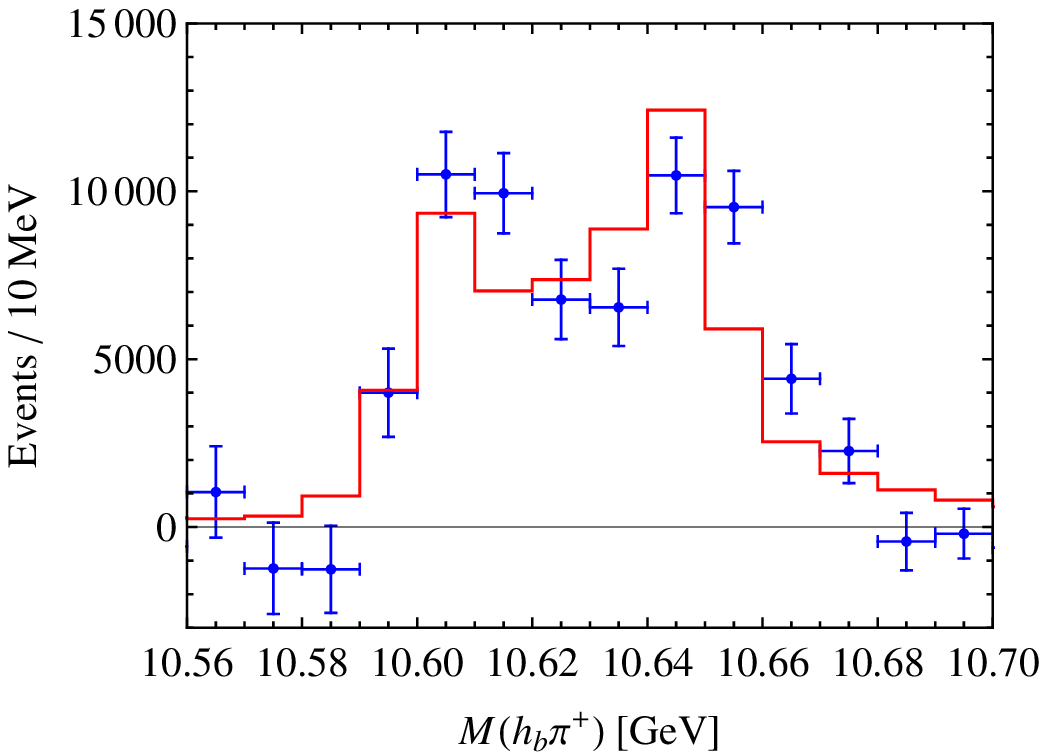}\\[5mm]
\includegraphics[width=0.48\textwidth]{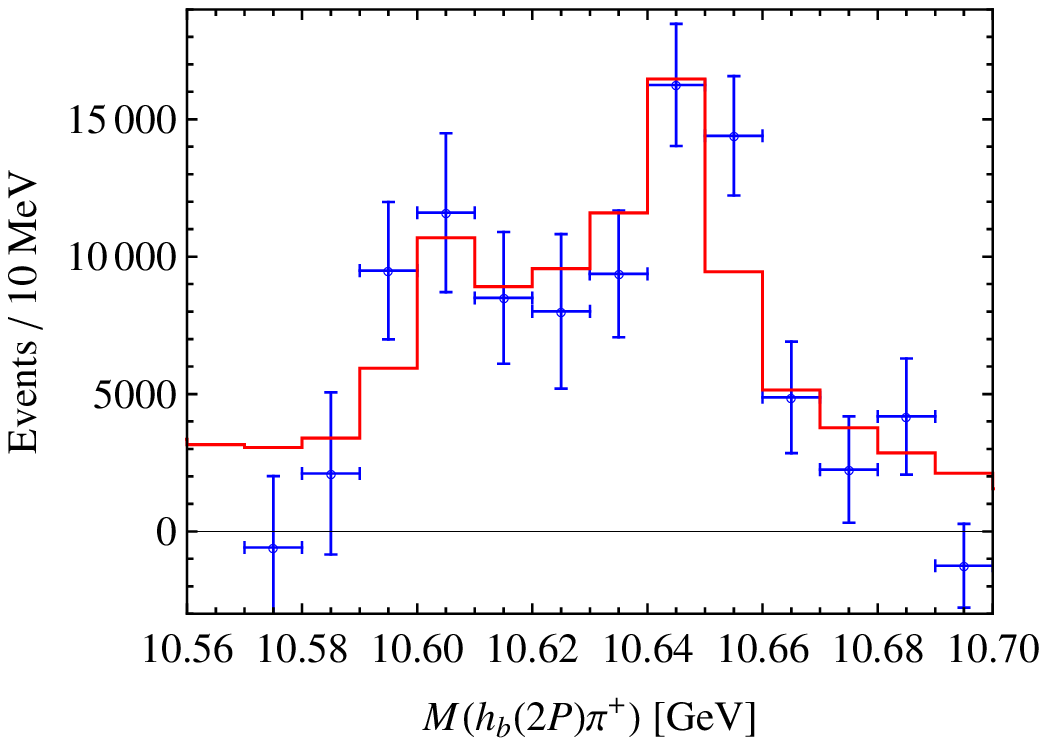}
\caption{Comparison of the calculated invariant mass spectra of $h_b\pi^+$ and $h_b(2P)\pi^+$ with
the measured missing mass spectra $MM(\pi)$. \label{fig:resbs}}
\end{center}
\end{figure}

In order to fit to the data which are events collected per 10~MeV, we integrate
the invariant mass spectra for each bin corresponding to the measurements. This
is important for narrow structures. The best fit results in $\chi^2/{\rm
d.o.f.}=54.1/22=2.45$.
\ba%
z_1 &=& 0.75^{+0.08}_{-0.11}~{\rm GeV}^{-1/2}, \qquad r_z = -0.39^{+0.06}_{-0.07},\non\\
g_1 z_1 &=& 0.40\pm0.06~{\rm GeV}^{-1}.
\ea%
The results from the best fit are plotted in Fig.~\ref{fig:resbs} together with
the experimental data. Using Eq.~(\ref{eq:zeff}) the
     couplings can be converted to binding energies. Especially we find
     \be%
     \Eng_{Z_b}= -4.7^{+2.2}_{-2.3}~{\rm MeV}, \qquad \Eng_{Z_b'}=
     -0.11^{+0.06}_{-0.14}~{\rm MeV} \ .
     \ee%

Although the $Z_b$ and $Z_b'$ are supposedly spin partners, a value of
$r_z=-0.4$ is not completely unreasonable: the fine tuning necessary to put a
bound state as close as 0.1~MeV to a threshold is extremely sensitive to even a
small variation in the scattering potential, driven by spin symmetry violations.
In effect this can give significant differences in the binding energies and, via
Eq.~(\ref{eq:zeff}), also in the coupling constants. However, a microscopic
calculation, which goes beyond the scope of this paper,
 would be necessary to check this hypothesis.

One can obtain a better fit to the data if one would either release the bound
state condition given as Eq.~(\ref{eq:zeff}), and allowing the masses of the
$Z_b$ states to float freely, or allow for non-resonant terms. But this is not
the purpose of our paper --- we here only want to demonstrate that the data are
consistent with the bound state picture, which implies the masses of the
$Z_b^{(\prime)}$ states to be located below the corresponding thresholds.

It is interesting to look at the $Z_b$ line shape or the absolute value of
$G_Z(E)$ for different locations of the $Z_b$ pole.
\begin{figure}[t]
\begin{center}
\includegraphics[width=0.435\textwidth]{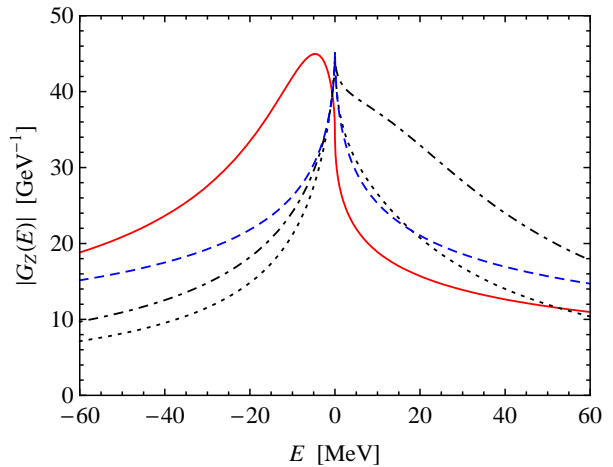}
\caption{The absolute value of $G_Z(E)$.
The solid (red) and dashed (blue) are for a bound state and virtual state, respectively,
with the same mass, $\Eng=M_{Z_b}-M_B-M_{B^*}=-4.7$~MeV. The
dotted and dot-dashed (black) lines are for resonances with $\Eng=8$ and 20~MeV, respectively.
The maxima of the resonance and virtual state curves have been normalized to the bound state one. \label{fig:gz}}
\end{center}
\end{figure}
The function $|G_Z(E)|$ using the parameters from the best fit are plotted as
the solid line in Fig.~\ref{fig:gz}. In this case, the $Z_b$ is a $B\bar B^*$
bound state with a binding energy of $-4.7$~MeV. Keeping $z_1=0.75~{\rm
  GeV}^{-1/2}$, and $g_1 z_1=0.4$~GeV$^{-1}$ fixed we also plot as the dashed
line the line shape for the virtual state with the same value of
$\Eng$.\footnote{The curve for virtual state is obtained using
  Eq.~(\ref{eq:gz0}) but with $\widetilde{\Sigma}(E)=\Sigma(E)-{\rm
    Re}(\Sigma_{\rm II}(\Eng))-(E-\Eng){\rm Re}(\Sigma'(\Eng)) \ ,$ where, in
  the $\overline{\rm MS}$ scheme, ${\rm Re}(\Sigma_{\rm II}(\Eng)) = -{\rm
    Re}(\Sigma(\Eng))$ is the self-energy in the second Riemann sheet.}
 The dotted and dot-dashed lines are for a
resonance with a mass above the $B\bar B^*$ threshold by 8 and 20~MeV,
respectively. From the figure, one sees that the bound state produces a bump
below the threshold, and a small cusp at the threshold, while the virtual state
produces a prominent cusp at the threshold and no structure below. Above the
threshold, the energy dependence of the virtual and bound state curves are
exactly the same. The bump in the bound state case reflects the pole position.
Hence, if we reduce the value of binding energy to, say about 0.1~MeV, which is
the case for the $Z_b'$ and the $X(3872)$, then the bump below threshold would
be invisible, and the cusp dominates the structure. In this case, it is hard to
distinguish between the bound state and virtual state scenarios. For more
discussions on the shape of a virtual state, see e.g.~\cite{mcvoy,Baru:2004xg}.

One important feature of the line shapes of dynamically generated states
 is shown in Fig.~\ref{fig:gz}: for poles slightly
above the threshold, since the coupling to the opening channel is strong,  the
position of the peak is \emph{locked to the threshold}, as can be seen from the
dotted line. Increasing the resonance mass, the effect of the cusp is smeared
out, and shape is approaching a normal Breit-Wigner resonance --- in the
dot-dashed line one starts to see a bump above threshold developing for a mass
as large as 20~MeV above the $B\bar B^*$ threshold. However, even then the peak
is still located at the threshold. We are therefore to conclude that a
Breit-Wigner parametrization, as was used in the experimental analysis, should
not be used when analyzing structures that emerge from dynamically generated
states.

\section{Conclusion}

We showed that the data~\cite{Adachi:2011gj} is consistent with the assumption
that the main components of the lower and higher $Z_b$ states are $S$-wave
$B\bar B^*+{\rm c.c.}$ and $B^*\bar B^*$ bound states, respectively. A small
compact tetraquark component, however, can not be excluded.

It is difficult to distinguish between resonance and bound state scenarios
with the current data, however, data with higher resolution should allow one
to distinguish the two cases --- see, e.g., Refs.~\cite{ourX,ericsX} for the
corresponding discussion for the $X(3872)$. In a next step we will improve the
model by inclusion of non-resonant terms as well as the calculation of
of other decay channels~\cite{future}.

We have demonstrated that if the  $Z_b$ states are indeed generated from
non-perturbative  $B\bar B^*+{\rm c.c.}$ and $B^*\bar B^*$ dynamics, the data
should not be analyzed using a Breit-Wigner parametrization. This statement can
also be reversed: if a near threshold state can be described by a Breit-Wigner
form, it is not dynamically generated, as this is possible only if the coupling
of the resonance to the continuum channel is very weak. At present the data
appears to be consistent with line shapes that result from dynamical states as
well as genuine ones. Therefore a decision about the nature of the $Z_b$ states
will be possible only once data with higher resolution and statistics will be
available.

\subsection*{Acknowledgments}
We thank the HGF for funds provided to the virtual institute ``Spin and strong
QCD'' (VH-VI-231), the DFG (SFB/TR 16) and the EU I3HP ``Study of Strongly
Interacting Matter'' under the Seventh Framework Program of the EU. U.-G. M.
also thanks the BMBF for support (Grant No. 06BN9006).

\begin{appendix}
\section{Loop function and expressions of the amplitudes}
\label{app:amp}
\renewcommand{\theequation}{\thesection.\arabic{equation}}
\setcounter{equation}{0}

The basic three-point loop function worked out using dimensional regularization
in $d=4$ is
\ba%
&& I(m_1,m_2,m_3,\vec{q})\non\\ &=& \frac{-i}{8} \int\!\frac{d^dl}{(2\pi)^d}
\frac{1}{ \left(l^0-\frac{\vec{l}^2}{m_1}+i\epsilon\right)}
 \frac{1}{\left(l^0+b_{12}+\frac{\vec{ l}^2}{m_2}-i\epsilon\right)} \non\\
 &&\times
 \frac{1}{\left[l^0+b_{12}-b_{23}-\frac{(\vec{l}-\vec{
q})^2}{m_3}+i\epsilon\right] } \non\\
&=& \frac{\mu_{12}\mu_{23}}{16\pi} \frac{1}{\sqrt{a}} \left[
\tan^{-1}\left(\frac{c'-c}{2\sqrt{a(c-i\epsilon)}}\right) \right.  \non\\
&& \left. + \tan^{-1}\left(\frac{2a+c-c'}{2\sqrt{a(c'-a-i\epsilon)}}\right)
\right],
\ea%
where $m_i(i=1,2,3)$ are the masses of the particles in the loop,
$\mu_{ij}=m_im_j/(m_i+m_j)$ are the reduced masses, $b_{12} = m_1+m_2-M$,
$b_{23}=m_2+m_3+q^0-M$ with $M$ the mass of the initial particle, and
$$
a = \left(\frac{\mu_{23}}{m_3}\right)^2 \vec{ q}^2, \quad c =
2\mu_{12}b_{12}, \quad c'=2\mu_{23}b_{23}+\frac{\mu_{23}}{m_3}\vec{ q}^2.
$$
For more information about the loop function, we refer to Appendix A in
Ref.~\cite{Guo:2010ak}. Note that different from the convention of
Ref.~\cite{Guo:2010ak}, here the factor $1/(m_1m_2m_3)$ has been dropped.

In terms of the loop function given above, the amplitudes for $Z_b^{+}$ and
$Z_b^{\prime+}$ decays into $h_b\pi^+$ are
\begin{align}
\Amp_{Z_b^{+}h_b} &= \frac{2\sqrt{2}g g_1 z_1}{F_\pi} \sqrt{M_{h_b}M_{Z_b}} \epsilon_{ijk}q^i
\varepsilon_{Z_b}^j \varepsilon_{h_b}^k \non\\
& \times \left[I(M_B,M_{B^*},M_{B^*},\vec{q})+I(M_{B^*},M_B,M_{B^*},\vec{q})\right], \label{Aeq:ahb1}\\
\intertext{and} %
\Amp_{Z_b^{\prime+}h_b} &= \frac{2\sqrt{2}g g_1 z_2}{F_\pi} \sqrt{M_{h_b}M_{Z_b'}}
\epsilon_{ijk}q^i \varepsilon_{Z_b'}^j \varepsilon_{h_b}^k \non\\
& \times
\left[I(M_{B^*},M_{B^*},M_B,\vec{q})+I(M_{B^*},M_{B^*},M_{B^*},\vec{q})\right], \label{Aeq:ahb2}
\end{align}
respectively. In all these amplitudes, both the neutral and charged bottom and
anti-bottom mesons have been taken into account.

\end{appendix}


\begin{thebibliography}{99}

\bibitem{Adachi:2011gj}
  I. Adachi et al. [Belle Collaboration],
  %``Observation of two charged bottomonium-like resonances,''
  arXiv:1105.4583 [hep-ex].
  %%CITATION = ARXIV:1105.4583;%%

%\cite{Bondar:2011ev}
\bibitem{Bondar:2011ev}
  A.~E.~Bondar, A.~Garmash, A.~I.~Milstein, R.~Mizuk and M.~B.~Voloshin,
  %``Heavy quark spin structure in Z_b resonances,''
  arXiv:1105.4473 [hep-ph].
  %%CITATION = ARXIV:1105.4473;%%


%\cite{Zhang:2011jj}
\bibitem{Zhang:2011jj}
  J.~R.~Zhang, M.~Zhong and M.~Q.~Huang,
  %``Could $Z_{b}(10610)$ be a $B^{*}\bar{B}$ molecular state?,''
  arXiv:1105.5472 [hep-ph].
  %%CITATION = ARXIV:1105.5472;%%


%\cite{Yang:2011rp}
\bibitem{Yang:2011rp}
  Y.~Yang, J.~Ping, C.~Deng and H.~S.~Zong,
  %``Dynamical study of the $Z_b$(10610) and $Z_b$(10650) as molecular states,''
  arXiv:1105.5935 [hep-ph].
  %%CITATION = ARXIV:1105.5935;%%


%\cite{Sun:2011uh}
\bibitem{Sun:2011uh}
  Z.~F.~Sun, J.~He, X.~Liu, Z.~G.~Luo and S.~L.~Zhu,
  %``$Z_b(10610)^\pm$ and $Z_b(10650)^\pm$ as the $B^*\bar{B}$ and
  %$B^*\bar{B}^{*}$ molecular states,''
  arXiv:1106.2968 [hep-ph].
  %%CITATION = ARXIV:1106.2968;%%


%\cite{Nieves:2011vw}
\bibitem{Nieves:2011vw}
  J.~Nieves and M.~P.~Valderrama,
  %``Deriving the existence of $B\bar{B}^*$ bound states from the X(3872) and
  %Heavy Quark Symmetry,''
  arXiv:1106.0600 [hep-ph].
  %%CITATION = ARXIV:1106.0600;%%


%\cite{Bugg:2011jr}
\bibitem{Bugg:2011jr}
  D.~V.~Bugg,
  %``An explanation of Belle states Z_a(10610) and Z_b(10650),''
  arXiv:1105.5492 [hep-ph].
  %%CITATION = ARXIV:1105.5492;%%

%\cite{Danilkin:2011sh}
\bibitem{Danilkin:2011sh}
  I.~V.~Danilkin, V.~D.~Orlovsky and Yu.~A.~Simonov,
  %``Hadron interaction with heavy quarkonia,''
  arXiv:1106.1552 [hep-ph].
  %%CITATION = ARXIV:1106.1552;%%

%\cite{Chen:2011pv}
\bibitem{Chen:2011pv}
  D.~Y.~Chen and X.~Liu,
  %``$Z_b(10610)$ and $Z_b(10650)$ structures produced by the initial single
  %pion emission in $\Upsilon(5S)$ decays,''
  arXiv:1106.3798 [hep-ph].
  %%CITATION = ARXIV:1106.3798;%%


%\cite{Voloshin:1982ij}
\bibitem{Voloshin:1982ij}
  M.~B.~Voloshin,
  %``ON A FOUR QUARK ISOVECTOR RESONANCE IN THE UPSILON FAMILY,''
  JETP Lett.\  {\bf 37}, 69 (1983)
  [Pisma Zh.\ Eksp.\ Teor.\ Fiz.\  {\bf 37}, 58 (1983)].
  %%CITATION = ZFPRA,37,58;%%

%\cite{Butler:1993rq}
\bibitem{Butler:1993rq}
  F.~Butler {\it et al.} [CLEO Collaboration],
  %``Analysis of hadronic transitions in upsilon (3S) decays,''
  Phys.\ Rev.\  D {\bf49}, 40(1994).


%\cite{Anisovich:1995zu}
\bibitem{Anisovich:1995zu}
  V.~V.~Anisovich, D.~V.~Bugg, A.~V.~Sarantsev and B.~S.~Zou,
  %``Upsilon (3S) ---> Upsilon (1S) pi pi decay: Is the pi pi spectrum puzzle an
  %indication of a b anti-b q anti-q resonance?,''
  Phys.\ Rev.\  D {\bf 51}, 4619 (1995).
  %%CITATION = PHRVA,D51,4619;%%

%\cite{Guo:2004dt}
\bibitem{Guo:2004dt}
  F.~K.~Guo, P.~N.~Shen, H.~C.~Chiang and R.~G.~Ping,
  %``Heavy quarkonium pi+ pi- transitions and a possible b anti-b q anti-q
  %state,''
  Nucl.\ Phys.\  A {\bf 761}, 269 (2005)
  [arXiv:hep-ph/0410204].
  %%CITATION = NUPHA,A761,269;%%
%
%\cite{Guo:2006ai}
\bibitem{Guo:2006ai}
  F.~K.~Guo, P.~N.~Shen, H.~C.~Chiang and R.~G.~Ping,
  %``On the structure of the pi pi invariant mass spectra of the Upsilon(4S) -->
  %Upsilon(1S, 2S) pi+ pi-,''
  Phys.\ Lett.\  B {\bf 658}, 27 (2007)
  [arXiv:hep-ph/0601120].
  %%CITATION = PHLTA,B658,27;%%

%\cite{Chen:2011zv}
\bibitem{Chen:2011zv}
  D.~Y.~Chen, X.~Liu and S.~L.~Zhu,
  %``Charged bottomonium-like states $Z_b(10610)$ and $Z_b(10650)$ and the
  %$\Upsilon(5S)\to \Upsilon(2S)\pi^+\pi^-$ decay,''
  arXiv:1105.5193 [hep-ph].
  %%CITATION = ARXIV:1105.5193;%%

%\cite{Flatte:1976xu}
\bibitem{Flatte:1976xu}
  S.~M.~Flatt\'{e},
  %``Coupled - Channel Analysis of the pi eta and K anti-K Systems Near K anti-K Threshold,''
  Phys.\ Lett.\  B {\bf63}, 224 (1976).

\bibitem{future}
  M.~Cleven, F.~K.~Guo, C.~Hanhart and U.-G.~Mei{\ss}ner, in preparation.


%\cite{Fleming:2008yn}
\bibitem{Fleming:2008yn}
  S.~Fleming, T.~Mehen,
  %``Hadronic Decays of the X(3872) to chi(cJ) in Effective Field Theory,''
  Phys.\ Rev.\ D  {\bf78}, 094019 (2008).
  [arXiv:0807.2674 [hep-ph]].

%\cite{Guo:2010ak}
\bibitem{Guo:2010ak}
  F.~K.~Guo, C.~Hanhart, G.~Li, U.-G.~Mei{\ss}ner and Q.~Zhao,
  %``Effect of charmed meson loops on charmonium transitions,''
  Phys.\ Rev.\  D {\bf 83}, 034013 (2011)
  [arXiv:1008.3632 [hep-ph]].
  %%CITATION = PHRVA,D83,034013;%%

%\cite{Burdman:1992gh}
\bibitem{Burdman:1992gh}
  G.~Burdman and J.~F.~Donoghue,
  %``Union of chiral and heavy quark symmetries,''
  Phys.\ Lett.\  B {\bf 280}, 287 (1992);
  %%CITATION = PHLTA,B280,287;%%
%
%%\cite{Wise:1992hn}
%\bibitem{Wise:1992hn}
  M.~B.~Wise,
  %``Chiral Perturbation Theory For Hadrons Containing A Heavy Quark,''
  Phys.\ Rev.\  D {\bf 45}, R2188 (1992);
  %%CITATION = PHRVA,D45,2188;%%
%
%\cite{Yan:1992gz}
%\bibitem{Yan:1992gz}
  T.~M.~Yan, H.~Y.~Cheng, C.~Y.~Cheung, G.~L.~Lin, Y.~C.~Lin and H.~L.~Yu,
  %``Heavy Quark Symmetry And Chiral Dynamics,''
  Phys.\ Rev.\  D {\bf 46}, 1148 (1992)
  [Erratum-ibid.\  D {\bf 55}, 5851 (1997)];
  %%CITATION = PHRVA,D46,1148;%%
%
%%\cite{Casalbuoni:1996pg}
%\bibitem{Casalbuoni:1996pg}
  R.~Casalbuoni, A.~Deandrea, N.~Di Bartolomeo, R.~Gatto, F.~Feruglio and G.~Nardulli,
  %``Phenomenology Of Heavy Meson Chiral Lagrangians,''
  Phys.\ Rept.\  {\bf 281}, 145 (1997)
  [arXiv:hep-ph/9605342].
  %%CITATION = PRPLC,281,145;%%

%\cite{Weinberg:1965zz}
\bibitem{molecule1}
  S.~Weinberg,
  %``Elementary particle theory of composite particles,''
  Phys.\ Rev.\  {\bf 130}, 776 (1963);
  %%CITATION = PHRVA,130,776;%%
  %S.~Weinberg,
  %``Quasiparticles and the Born Series,''
  Phys.\ Rev.\  {\bf 131}, 440 (1963);
  %%CITATION = PHRVA,131,440;%%
  %S.~Weinberg,
  %``Evidence That the Deuteron Is Not an Elementary Particle,''
  Phys.\ Rev.\  {\bf 137}, B672 (1965);
  %%CITATION = PHRVA,137,B672;%%
%\bibitem{molecule2}
%\cite{Baru:2003qq}
%\bibitem{Baru:2003qq}
  V.~Baru, J.~Haidenbauer, C.~Hanhart, Yu.~Kalashnikova and A.~E.~Kudryavtsev,
  %``Evidence that the a0(980) and f0(980) are not elementary particles,''
  Phys.\ Lett.\  B {\bf 586}, 53 (2004)
  [arXiv:hep-ph/0308129].
  %%CITATION = PHLTA,B586,53;%%

%\cite{Guo:2009wr}
\bibitem{Guo:2009wr}
  F.~K.~Guo, C.~Hanhart and U.-G.~Mei{\ss}ner,
  %``On the extraction of the light quark mass ratio from the decays psi-prime
  %---> J/psi pi0 (eta),''
  Phys.\ Rev.\ Lett.\  {\bf 103}, 082003 (2009)
  [Erratum-ibid.\  {\bf 104}, 109901 (2010)]
  [arXiv:0907.0521 [hep-ph]].
  %%CITATION = PRLTA,103,082003;%%

%\cite{Guo:2010zk}
\bibitem{Guo:2010zk}
  F.~K.~Guo, C.~Hanhart, G.~Li, U.-G.~Mei{\ss}ner and Q.~Zhao,
  %``Novel analysis of the decays psi' -> h_c pi^0 and eta_c'-> chi_{c0} pi^0,''
  Phys.\ Rev.\  D {\bf 82}, 034025 (2010)
  [arXiv:1002.2712 [hep-ph]].
  %%CITATION = PHRVA,D82,034025;%%

\vfill\eject

\bibitem{PDG2010}
 K.~Nakamura {\it et al.} [Particle Data Group], J.\ Phys.\ G {\bf 37}, 075021 (2010).

%\cite{Li:2010rh}
\bibitem{Li:2010rh}
  X.~Q.~Li, F.~Su and Y.~D.~Yang,
  %``Determination of the strong coupling $g_{B^* B\pi}$ from semi-leptonic
  %$B\to \pi \ell \nu$ decay,''
  Phys.\ Rev.\  D {\bf 83}, 054019 (2011)
  [arXiv:1011.0269 [hep-ph]].
  %%CITATION = PHRVA,D83,054019;%%

%\cite{Guo:2009ct}
\bibitem{Guo:2009ct}
  F.~K.~Guo, C.~Hanhart and U.-G.~Mei{\ss}ner,
  %``Interactions between heavy mesons and Goldstone bosons from chiral
  %dynamics,''
  Eur.\ Phys.\ J.\  A {\bf 40}, 171 (2009)
  [arXiv:0901.1597 [hep-ph]].
  %%CITATION = EPHJA,A40,171;%%

\bibitem{nn2dpi}
 V.~Lensky, V.~Baru, J.~Haidenbauer, C.~Hanhart, A.~E.~Kudryavtsev, U.-G.~Mei{\ss}ner,
  %``Towards a field theoretic understanding of NN ---> NN pi,''
  Eur.\ Phys.\ J.\  A {\bf27}, 37 (2006)
  [nucl-th/0511054].

\bibitem{mcvoy}
  K.~W.~McVoy, Nucl.\ Phys.\ {\bf A115}, 481   (1968).

%\cite{Baru:2004xg}
\bibitem{Baru:2004xg}
  V.~Baru, J.~Haidenbauer, C.~Hanhart, A.~E.~Kudryavtsev, U.-G.~Mei{\ss}ner,
  %``Flatte-like distributions and the a(0)(980) / f(0)(980) mesons,''
  Eur.\ Phys.\ J.\  A {\bf23}, 523 (2005)
  [nucl-th/0410099].


\bibitem{ourX}
  C.~Hanhart, Y.~S.~Kalashnikova, A.~E.~Kudryavtsev, A.~V.~Nefediev,
  %``Reconciling the X(3872) with the near-threshold enhancement in the D0 anti-D*0 final state,''
  Phys.\ Rev.\ D  {\bf 76},  034007 (2007)
  [arXiv:0704.0605 [hep-ph]].

\bibitem{ericsX}
  E.~Braaten, M.~Lu,
  %``Line shapes of the X(3872),''
  Phys.\ Rev.\ D {\bf 76}, 094028 (2007)
  [arXiv:0709.2697 [hep-ph]].



\end{thebibliography}
\end{document}